\documentclass[preprint,authoryear,3p]{elsarticle}
\usepackage{lineno,hyperref}
\usepackage{mathtools}
\usepackage{multirow}
\usepackage{multicol}
\usepackage{booktabs}
\usepackage{listings}
\usepackage{xcolor}
\usepackage{footmisc}
\usepackage[font=normalsize]{caption}
\usepackage[ruled,vlined]{algorithm2e}

\modulolinenumbers[5]

\journal{Expert Systems with Applications}
\biboptions{authoryear}

\makeatletter
\g@addto@macro{\UrlBreaks}{\UrlOrds}
\makeatother

\usepackage{blindtext}
\usepackage{hyperref}
\usepackage{nameref}

\begin{document}

\begin{frontmatter}

\title{Targeted aspect-based emotion analysis to detect opportunities and precaution in financial Twitter messages}

\author[mymainaddress]{Silvia Garc\'ia-M\'endez\corref{mycorrespondingauthor}}
\ead{sgarcia@gti.uvigo.es}
\author[mymainaddress]{Francisco de Arriba-P\'erez}
\ead{farriba@gti.uvigo.es}
\author[mymainaddress]{Ana Barros-Vila}
\ead{abarros@gti.uvigo.es}
\author[mymainaddress]{Francisco J. Gonz\'alez-Casta\~no}
\ead{javier@det.uvigo.es}
\address[mymainaddress]{Information Technologies Group, atlanTTic, University of Vigo, EI Telecomunicaci\'on, Campus Lagoas-Marcosende, Vigo, 36310, Spain}

\cortext[mycorrespondingauthor]{Corresponding author: sgarcia@gti.uvigo.es}

\begin{abstract}
Microblogging platforms, of which Twitter is a representative example, are valuable information sources for market screening and financial models. In them, users voluntarily provide relevant information, including educated knowledge on investments, reacting to the state of the stock markets in real-time and, often, influencing this state. We are interested in the user forecasts in financial, social media messages expressing opportunities and precautions about assets. We propose a novel Targeted Aspect-Based Emotion Analysis (\textsc{tabea}) system that can individually discern the financial emotions (positive and negative forecasts) on the different stock market assets in the same tweet (instead of making an overall guess about that whole tweet). It is based on Natural Language Processing (\textsc{nlp}) techniques and Machine Learning streaming algorithms. The system comprises a constituency parsing module for parsing the tweets and splitting them into simpler declarative clauses; an offline data processing module to engineer textual, numerical and categorical features and analyse and select them based on their relevance; and a stream classification module to continuously process tweets on-the-fly. Experimental results on a labelled data set endorse our solution. It achieves over 90\% precision for the target emotions, financial opportunity, and precaution on Twitter. To the best of our knowledge, no prior work in the literature has addressed this problem despite its practical interest in decision-making, and we are not aware of any previous \textsc{nlp} nor online Machine Learning approaches to \textsc{tabea}.

\end{abstract}


\end{frontmatter}

\section{Introduction}

In this section, the context of the work, as well as the research problem, will be discussed, paying special attention to sentiment and emotion analysis fields towards targeted aspect-based emotion analysis, in which the proposed solution is framed. Then, the contributions of the research will be described along with the paper organisation.

\subsection{Application context}

Crowdsourcing in Web 2.0 and beyond and, particularly, the rapid development of social media platforms, have produced massive digital content in many sectors, including finance \citep{Sohangir2018}. Microblogging and social trading networks allow users to track the stock markets' behaviour from the valuable comments by expert users and other screening data. These sources are widely used by financial screening solutions owing to the real-time data provided. They are useful to generate indicators on asset pricing \citep{Houlihan2021}, loan and insurance underwriting \citep{Bee2021}, and financial risk prevention \citep{Gao2021,Li2021} for novice traders and stockholders \citep{Ge2020}.

It has been reported that user contributions to social media platforms such as Twitter\footnote{Available at \url{https://twitter.com}, January 2023.} clearly influence the behaviour of stock markets \citep{Li2018,Mai2018,Ronaghi2022}. This type of content has been analysed to successfully predict sales \citep{Pai2018,Yuan2018,Kim2021}. Some social trading platforms such as eToro\footnote{Available at \url{https://www.etoro.com}, January 2023.} and \textsc{xtb}\footnote{Available at \url{https://www.xtb.com}, January 2023.} also keep users' reviews and comments along with past pricing data on the assets they trade, which can be processed. For example, the Thomson Reuters Eikon platform\footnote{Available at \url{https://eikon.thomsonreuters.com}, January 2023.} claims to analyse financial news and market sentiments.

\subsection{Research problem}

The user opinions on the web pages and social platforms can only be analysed with automatic methods \citep{Kang2018,Hemmatian2019,Dau2020,Alamoudi2021,Nilashi2021,Wang2022} based on Artificial Intelligence (\textsc{ai}) techniques such as Natural Language Processing (\textsc{nlp}) and Machine Learning, due to the noise and complexity of the messages.

From an application perspective, our work pursues tools used to detect market speculations in social media posts, whether expressing opportunities or precautions, about the several financial assets or stock markets that may be present in the same social media post, to support decision-making in finance \citep{Li2020}.

From an analytical perspective, the problem in this work is modelled as a Targeted Aspect-Based Emotion Analysis (\textsc{tabea}), which derives from the homologous problem in sentiment analysis and is related to Opinion Target Extraction (\textsc{ote}). There exist two approaches within Aspect-Based Sentiment Analysis (\textsc{absa}): Aspect-Category Sentiment Analysis (\textsc{acsa}) \citep{Liao2021} and Aspect-Term Sentiment Analysis (\textsc{atsa}) \citep{Peng2022}. \textsc{acsa} seeks to extract the polarity of predefined aspect categories even though most of the time these aspects are not explicit in the text (\textit{e.g.}, ``they offer high-quality steaks'', referring to aspect category ``food'' of implicit target ``hotel''). \textsc{atsa} extracts aspects related to certain target aspects/entities (for example, ``Hilton''). When \textsc{absa} considers several different entities, it is referred to as Targeted Aspect-Based Sentiment Analysis (\textsc{tabsa}). The only difference with \textsc{tabea} is that our work focuses on emotions \citep{Chen2018,Rout2018,Polignano2021,Shahin2022}, rather than on polarities, although it is also applied at aspect/entity levels. This is necessary because financial news or social media posts often contain different aspects and emotions, and we seek an accurate classification system conducting fine-grained emotion analysis of stock market assets.

\subsection{From basic sentiment and emotion analysis to online \textsc{tabea}}

Sentiment and emotion analysis in their more general forms, whether applied at a coarse or fine-grained level to finance, are relevant to our research problem. They have inspired us to take certain practical considerations as the Machine Learning approach selected.

Sentiment analysis of user opinions has been a prolific research field \citep{Madhu2018,Nurifan2019,Yue2019}. Nowadays, sentiment analysis can automatically process large-scale data from social media platforms \citep{Yue2019}, blogs \citep{Madhu2018}, online communities and fora such as wikis \citep{Nurifan2019} and other collaborative media. Sentiment analysis often relies on \textsc{nlp} techniques \citep{Dogra2021}, mostly on English texts. It usually classifies texts into three polarity levels (negative, neutral, and positive) and can be performed at document \citep{Rhanoui2019}, sentence \citep{Arulmurugan2019} and aspect/entity \citep{Mowlaei2020} levels. Both coarse and fine-grained sentiment analysis approaches have traditionally relied on Machine Learning algorithms whose accuracy depends on the availability of large labelled data sets with enough context information for the target domain \citep{Elnagar2018}. Document-oriented sentiment analysis seeks an overall document polarity score, regardless of the fact that documents are composed of many sentences where distinct targets and polarities coexist. The same applies to the sentence level. Different topics (asset {\it tickers} in this research) in a sentence may have different respective polarities. Depending on the depth of the level, sentiment analysis may be more complex. Coarse-grained document and even sentence-level sentiment analysis cannot handle the different aspects of users' comments and their associated polarities. This is the goal of entity/aspect-based sentiment analysis \citep{Ozyurt2021}, which focuses on the polarities of users' opinions about specific features (aspects) of the products or services (topics) under evaluation. 

Previous work on financial emotion analysis to characterise the behaviour of stock markets, on which our research focuses, is scarce \citep{DeArriba-Perez2020,Duxbury2020}. In fact, most previous works have exclusively focused on human emotions such as anger, fear, joy, love, sadness, and surprise \citep{Nandwani2021}. Nowadays, the interest in emotion analysis is widely recognised in academic and industrial research. Thus, we contribute to the solution of this problem with the particular approach of \textsc{tabea}, as previously stated.

At a processing framework level, online or streaming Machine Learning \citep{Benllarch2021,Mohawesh2021,Seth2021} is useful to handle real-time data in business applications \citep{Baier2020}, compared to Machine Learning batch approaches that handle changing input datasets with overall periodical retraining, even if this involves outdated data \citep{Pugliese2021}. In finance, unlike batch classification, online classification allows discovering relevant events as soon as they occur and taking early decisions \citep{Ko2021,Ko2022}.

\subsection{Contributions}

We present a Machine Learning system for \textsc{tabea} supported by \textsc{nlp} techniques specifically designed to detect two new financial emotions on textual content of the widely used Twitter platform, ``opportunity'' \citep{DeArriba-Perez2020} and ``precaution''. In the terminology by \cite{Plutchik2004}, “opportunity” would roughly correspond to positive emotions such as ``expectation” or ``anticipation”. ``Precaution” would be related to ``disapproval”, a negative opinion and, in a financial environment, it could correspond to a pessimistic feeling about a company or asset. These two new emotions allow identifying tweets that speculate about asset rises and falls. 

To the best of our knowledge, no prior work has applied \textsc{nlp} to \textsc{tabea} in streaming mode, either in finance or any other field. Thus, our problem and its analytical solution are the original contributions of this work. Besides, with the exception of our preliminary research on opportunity detection \citep{DeArriba-Perez2020}, the two new emotions we consider are also novel as here considered.

\subsection{Paper organisation}

The rest of this work is organised as follows. Section \ref{sec:related_work} discusses related work on Opinion Mining, sentiment analysis and emotion analysis from different perspectives (the use of \textsc{nlp} techniques, Machine Learning algorithms, etc.). Section \ref{sec:system} presents our classification problem and the architecture of our solution. Section \ref{sec:results} provides details about the financial data set and its features, and validates our approach with experimental results. Finally, Section \ref{sec:conclusions} concludes the paper.

\section{Related work}
\label{sec:related_work}

As previously mentioned, sentiment and emotion analysis related works will be discussed in this section, with special consideration to their application in the financial domain. Finally, explainability and data analysis in streaming will be discussed as relevant related fields.

\subsection{Sentiment analysis}

Researchers use techniques such as feature-engineering to characterise textual content like user comments \citep{Carrillo-de-Albornoz2018,Nawangsari2019}. 
In this regard, \cite{Weichselbraun2017} used an aspect lexicon that was built from training corpora, including ConceptNet\footnote{Available at \url{https://conceptnet.io}, January 2023.}, Wikipedia\footnote{Available at \url{https://es.wikipedia.org}, January 2023.} and Wordnet\footnote{Available at \url{https://wordnet.princeton.edu/}, January 2023.} to identify aspects. The more sophisticated solution by \cite{Liu2021}, a Global Semantic Memory Network for \textsc{absa}, enriches the representation of the targeted aspects with context information. Polarity lexica such as WordNet \citep{Jimenez2019} and SentiWordNet \citep{Madani2020} allow extracting the most likely sentiment that is associated with a certain word \citep{Mumtaz2018}. Then, sentiment propagation methods can be applied to provide a final score \citep{Li2018propagation}. For example, \cite{Fernandez-Gavilanes2018CreatingDescriptions} presented an unsupervised Machine Learning algorithm to automatically generate sentiment lexica from sentiment scores that were propagated across syntactic trees. Nevertheless, unsupervised approaches have been exclusively applied to coarse-grained document-level sentiment analysis so far. A promising strategy both for supervised and unsupervised approaches is stacking, which consists in combining different Machine Learning techniques to improve performance \citep{Mehmood2018,Wang2019} and allows handling neutrality and sentiment ambivalence \citep{Valdivia2018,Macdonald2019,DeArriba-Perez2020} by filtering neutral opinions to improve the performance of binary polarity classification. Moreover, \cite{Wang2020} presented a multi-level fine-scaled sentiment analysis system with ambivalence handling at the sentence level using keywords like \textit{extremely} and \textit{minor}, and then transferring polarity to paragraph/article levels with aggregation methods (\textit{i.e.}, sum function) and manual rules.

Sentiment analysis research has recently begun to apply graph neural models such as Convolutional Neural Networks \citep{Liang2022,Zhao2022}. Ongoing research also exploits syntactic and semantic structures into graph neural networks \citep{An2022,Phan2022,Xiao2022}. For example, \cite{Xiao2022} employed information of syntactic dependency trees from part-of-speech (\textsc{pos}) graphs. Then, contextual semantic dependencies are extracted using a syntactic distance attention mechanism over a densely connected graph convolutional network. In \cite{Liang2022}, a graph convolutional network considers sentence-aspect affective dependencies extracted from SenticNet\footnote{Available at \url{https://sentic.net}, January 2023.}. Moreover, the heterogeneous graph neural network by \cite{An2022} is composed of the word, aspect, and sentence nodes. The resulting structure and semantic data allow updating feature embeddings. \cite{Zhao2022} presented an aggregated graph convolutional network with an attention mechanism to model sentiment dependencies. \cite{Phan2022} exploited syntax-, semantic-, and context-based graphs as part of a convolutional neural network with an attention mechanism for aspect-level sentiment analysis.

\subsection{Emotion analysis}

\cite{Matsumoto2022} have performed an emotional breakdown analysis for chatbot solutions. They detect emotions in users' utterances with deep neural networks with sentence representation vectors. \cite{Guo2022} has also developed a deep learning-based solution by applying semantic text analysis for human emotion detection. Thus, no domain-focused emotions have been defined for finance, with the exception of our previous research on opportunity detection \citep{DeArriba-Perez2020}. For example, \cite{Razi2017} presented a multi-layer perceptron to analyse the influence of the aforementioned ``general'' emotions on investing behaviour. This was also the case of \cite{Zhou2018}, who studied the correlation between the trends of the Chinese stock market and five relevant features derived from ``general'' emotions. Other related problems can be found in \cite{Taffler2018,Chung2020,Ahn2021}.

\cite{Yuan2018mining} proposed a novel Emotion Latent Dirichlet Allocation (\textsc{elda}) model for credit rating prediction from Twitter data. \cite{Akhtar2020} presented a stacked ensemble Machine Learning solution to predict sentiments and emotions, but they only considered sentiment analysis nor emotion analysis in the financial domain. \textsc{aspendys} \citep{Theodorou2021} is an \textsc{ai} investment platform that mines financial text from Twitter and Stocktwits\footnote{Available at \url{https://stocktwits.com}, January 2023.} searches to assist investors with personalised recommendations. However, it is only based on sentiment analysis. \cite{Chun2021} employed a simple Emotion Term Frequency–Inverse Emotion Document Frequency (\textsc{etf}- \textsc{iedf}) technique derived from the classical Term Frequency-Inverse Document Frequency (\textsc{tf}- \textsc{idf}) \citep{Qaiser2018} instead of sentiment analysis to predict stock market prices using Machine Learning models, with good results. Finally, \cite{Valle-Cruz2022} analysed the influence of stock markets on investors emotions.

\subsection{Financial domain}

The knowledge extraction system by \cite{Wang2017} analysed sentiment time series from microblogs for stock market forecasting. \cite{Dridi2019} proposed a supervised Machine Learning sentiment analysis model for finance with competitive performance based on semantic features. \cite{Khan2020} proved that public sentiment and the political context affect stock market trends. They studied data from Yahoo! Finance\footnote{Available at \url{https://finance.yahoo.com}, January 2023.}, Twitter messages and Wikipedia political information to differentiate between stock markets that are hardly predictable and those that are easily influenced by financial news and social media. They combined spam filtering and feature selection with ensemble classifiers. The more sophisticated solution by \cite{Dogra2021} applies deep contextual language representation into the DistilBERT supervised sentiment analysis model \citep{Hew2020}.

In our work, based on the fact that a given financial text from the news or social media posts often contains different aspects and polarities, we seek an accurate classification system conducting fine-grained emotion analysis of stock market assets. As previously said, no previous work has considered financial opportunity and precaution from a \textsc{tabea} perspective.

\subsection{Interpretability and explainability}

Few recent state-of-the-art \textsc{ai} solutions have explored the interpretability and explainability of the models to explain their outcome to the end users to foster trustworthiness \citep{Gaur2021}. They apply symbolic techniques \citep{cambria2022senticnet} (\textit{e.g.}, auto-regressive language models, counterfactual explanations, kernel methods, post-hoc interpretability, rule-based explanations and statistical methods).

\subsection{Data streaming analysis}

To avoid the disadvantages of batch retraining \citep{Daniel2016,Turchi2017}, there exist real-time data streaming analysis alternatives, such as Massive Online Analysis (\textsc{moa})\footnote{Available at \url{https://moa.cms.waikato.ac.nz}, January 2023.}, Scikit-multiflow\footnote{Available at \url{https://scikit-multiflow.readthedocs.io/}, January 2023.} and StreamDM\footnote{Available at \url{http://huawei-noah.github.io/streamDM}, January 2023.}. These methods are appropriate for the fast, continuous and imbalanced real-time data from dynamic environments like social media platforms. They also allow for analysing the temporal evolution of input data. However, they have been largely unexplored in financial use cases, with some exceptions like the work by \cite{Nagarajan2019}. They describe an online sentiment analysis system for Twitter messages that outperforms traditional batch Machine Learning classifiers. However, they do not consider emotions, either financial or any other.

\subsection{Summary}
 
Summing up, no previous authors have considered \textsc{tabea} of financial, social media to detect investment opportunities or precautions about financial assets. In fact, as previously said, no prior work has applied \textsc{nlp} nor online Machine Learning to \textsc{tabea}, either in finance or any other field.

\section{System architecture}
\label{sec:system}

\begin{figure*}[!htbp]
 \centering
 \includegraphics[scale=0.18]{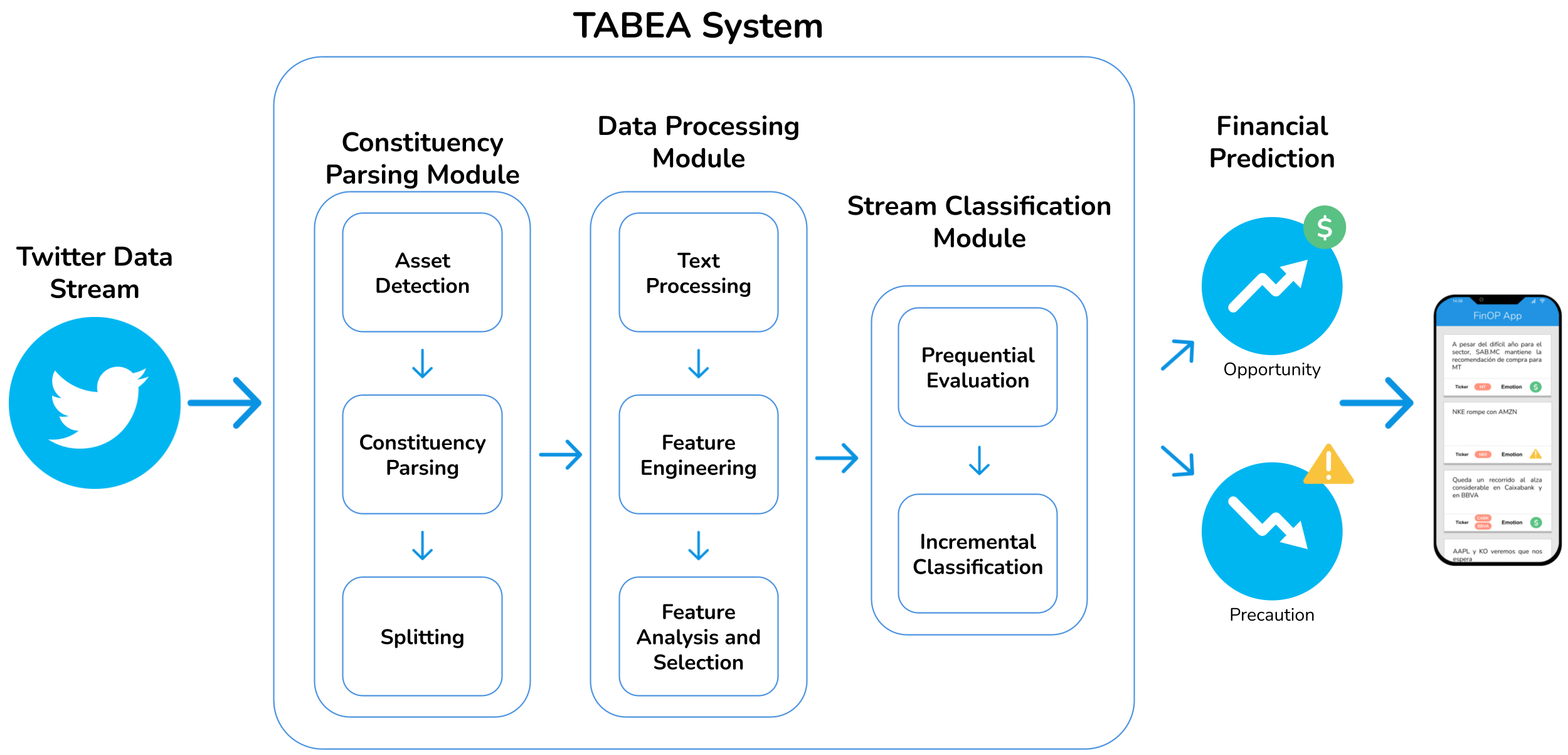}
 \caption{System scheme.}
 \label{fig:architecture}
\end{figure*}

In the following sections, we describe in detail our novel \textsc{tabea} system for fine-grained detection of financial opportunities and precautions on Twitter messages about stock markets and assets. Figure \ref{fig:architecture} shows the system scheme. Unlike our previous work \citep{DeArriba-Perez2020} as well as other prior work in the literature, it is a streaming scheme handling continuous flows of information and thus facilitating retraining of Machine Learning models. It is composed of (\textit{i}) a constituency parsing module for tweet parsing and splitting; (\textit{ii}) an offline data processing module to engineer features and select the best ones based on their relevance; and (\textit{iii}) a stream classification module that processes tweets on-the-fly.

\subsection{Constituency parsing module}
\label{sec:constituency_parsing_module_architecture}

\begin{figure}[htbp!]
 \centering
 \includegraphics[scale=0.10]{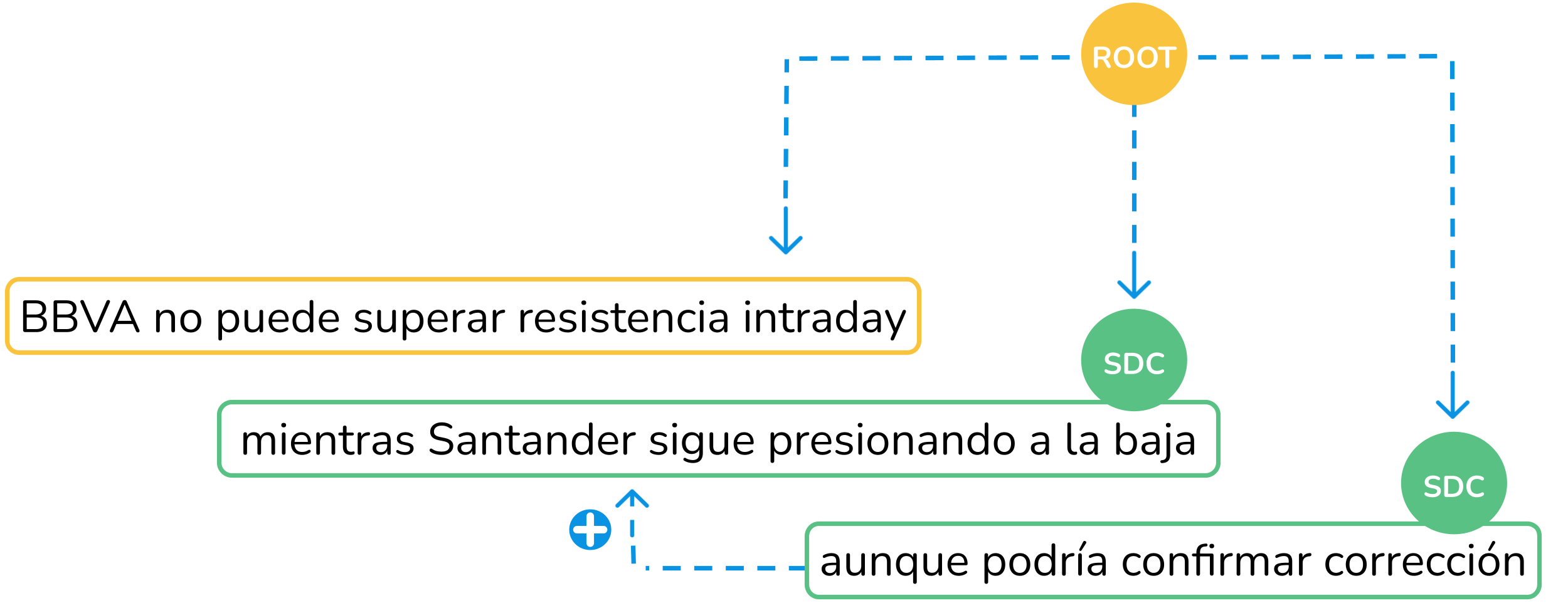}
 \caption{Tweet segments after applying forward propagation.}
 \label{fig:coreNLP}
\end{figure}

This module supports \textsc{tabea} through constituency parsing (tweet structure tree) and boundary splitting based on asset detection and linguistic features (morphology and syntax). Particularly, we detect Simple Declarative Clauses or \textsc{sdc} (\textit{i.e.}, those not introduced by a subordinating conjunction or a wh-word, and without subject-verb inversion) within the textual content of the tweets. The outcome of the module is a hierarchical dependency graph of the words that compose the tweet, as shown in Figure \ref{fig:coreNLP}.

Firstly, all assets in the tweet are identified using our financial lexica\footnote{\label{financial_lexica}Available at \url{https://www.gti.uvigo.es/index.php/en/resources/14-resources-for-finance-knowledge-extraction}, January 2023.}, and the tweets are split into simple declarative clauses. Then, the resulting segments are grouped if the following rules hold and forward propagated otherwise:

\begin{enumerate}
 \item The next segment does not contain any asset nor the additive conjunction \textit{y} `and' nor a comma (,) nor a hyphen (-).
 \item The next segment is a relative clause, \textit{i.e.}, it starts with 
 the relative conjunction \textit{que} `that' followed by a syntagm that contains any asset.

\end{enumerate}

In the example in Figure \ref{fig:coreNLP}, the two green clauses are grouped due to condition 1. The final segments after applying forward propagation are: \textit{\textsc{bbva} no puede superar resistencia intraday} `\textsc{bbva} cannot overcome intraday resistance' and \textit{mientras Santander sigue presionando a la baja aunque podría confirmar corrección} `while Santander continues to press down although it could confirm a correction'.

After the grouping is completed, a final segmentation based on regular expressions is performed to address the specific case of tweets containing lists of several assets (\textit{e.g.}, \textit{\textsc{alua.ba} -2,57\% \textsc{edn} +8,08\% \textsc{cres.ba} -4,86\%...}). In this particular and rather typical scenario, segments are split again right where the next asset is detected.

In the end, only segments containing at least one asset remain, and the other segments are discarded by the constituency parsing module. Algorithm \ref{alg:segmentation} summarises the actions performed by the module.

\SetKwComment{Comment}{/* }{*/}
\begin{algorithm*}[ht!]
 \caption{Constituency parsing and forward propagation}\label{alg:segmentation}
 \DontPrintSemicolon
 \KwIn{Financial lexica and textual content from the tweet}
 \KwData{financial\_lexica, tweet\_text, numbers (regular expression to detect numerical characters)}
 \KwResult{List of segments with a single asset or few strongly related assets}
\Begin{

 $tweet\_segmented$ = \textbf{declarative\_clause\_segmenter}($tweet\_text$);\Comment*[r]{List of segments from the tweet text obtained by constituency parsing. Assets are identified by the \textsc{ticker} tag.}
 
 $segment\_grouped$ = []
 
 $segment\_aux$=``''

 \ForEach(){$segment$ \textbf{in} $tweet\_segmented$}{

 \uIf{\textsc{ticker} \textbf{\textsc{or}} ``and" \textbf{\textsc{or}} ``," \textbf{\textsc{or}} ``-" \textbf{in} $segment$} {
 
 $segment\_grouped$.\textbf{add}($segment\_aux$)
 
 $segment\_aux$=$segment$
 }\uElse{
 $segment\_aux$+=$segment$ 
 \Comment*[r]{Group segments based on consideration 1.}
 }
 }
 $segment\_grouped$.\textbf{add}($segment\_aux)$
 
 $segment\_grouped.\textbf{remove\_empty\_segments}()$
 
 $list\_segments$= $segment\_grouped.\textbf{copy}()$
 
 $segment\_grouped$ = []
 
 \ForEach(){$r,\; r+1$ \textbf{in} \textbf{range}(\textbf{len}($list\_segments$)-1)}{
 \uIf{\textsc{ticker} \textbf{in} $list\_segments[r]$ \textbf{and} \textsc{ticker} \textbf{in} $list\_segments[r+1]$ \textbf{and} $list\_segments[r+1]$ \textbf{starts\_with} ``that"} {

 $list\_segments$[r+1]=($list\_segments[r]+list\_segments[r+1]$)
 \Comment*[r]{Group segments based on consideration 2.}

 }\uElse{
 $segment\_grouped$.\textbf{add}($list\_segments[r]$)
 }
 $segment\_grouped$.\textbf{add}($list\_segments[\textbf{len}(list\_segments)-1]$)

 }
 $result$ = []
 
 \ForEach(){$segment$ \textbf{in} $segment\_grouped$}{

 \uIf{$segment$.\textbf{regex}(``{\normalfont numbers}")$>$1 \Comment*[r]{Regular expression to detect numerical characters.}}
 {
 $result$.\textbf{addAll}($regex$.\textbf{split\_when\_found}($segment$,\textsc{ticker}))
 \Comment*[r]{To address the specific case of those tweets reporting the state of several assets.}
 }\uElse{$result$.\textbf{add}($segment$)} 
 }

 }
 
 \tcp*[h]{Returns the tweet segments} \; 
 $result.\textbf{keep\_segments\_with\_ticker}()$

 \textbf{return} $result$ \;
\end{algorithm*}

\subsection{Data processing module}
\label{sec:data_processing_architecture}

This module processes the text, generates and engineers features and selects the most relevant ones for classification.

\subsubsection{Text processing}
\label{sec:text_processing_architecture}
Text processing improves the efficiency of classification. It comprises the detection of asset and numeric values, filtering, cleaning and removing unnecessary information; hashtag splitting; and text lemmatisation tasks.

\begin{itemize}

\item \textbf{Asset identification}: we tag all financial assets with the \textsc{ticker} tag (note that this is consistent with the fact that the rules in Algorithm \ref{alg:segmentation} refer to any asset). For this purpose, we use the same dictionaries mentioned in Section \ref{sec:constituency_parsing_module_architecture}\footref{financial_lexica}. Moreover, numbers and percentages are replaced by \textsc{negative} or \textsc{positive} tags, depending on the mathematical signs that precede them, and, otherwise, by the \textsc{number} tag.

\item \textbf{Filtering, cleaning and removal}: we detect and remove symbols \$, @ and \#. Meaningless words such as connectors\footnote{Available at \url{https://www.ranks.nl/stopwords/spanish}, January 2023.} are also removed from the text along with \textsc{url}s and retweet (\textsc{rt}) tags. We keep words \textit{no} `not', \textit{sí} `yes', \textit{muy} `very' and \textit{poco} `few' due to their semantic load since they provide nuances of assets' trends.
 
\item \textbf{Hashtag splitting}: in order to maximise relevant textual data in the tweets and to fix spelling mistakes of the users, we use our lexica \citep{Garcia-Mendez2018,Garcia-Mendez2019} and the Spanish \textsc{crea} frequency reference corpus by {\it Real Academia Espa\~nola de la Lengua}\footnote{Available at \url{http://corpus.rae.es/lfrecuencias.html}, January 2023.} to decompose hashtags and compounds such as {\it mayorca\'ida} `biggestfall' into {\it mayor caída} `biggest fall'.
 
\item \textbf{Text lemmatisation}: the textual content is first tokenised into words and checked using a Spanish dictionary to keep only correct words. Finally, the tokens are lemmatised, and spelling mistakes are corrected with our text distance algorithm by replacing them with the most likely candidate (using again the \textsc{crea} corpus).

\end{itemize}

Table \ref{text_processed_tweets} illustrates the operation of the text processing stage.

\begin{table*}[!htbp]
\centering
\caption{\label{text_processed_tweets} Example of tweet after text processing.}
\begin{tabular}{cc}
\toprule
& \textbf{Tweet} \\\hline
\multirow{3}{*}{\bf Before} & 
\begin{tabular}[c]{@{}p{14cm}@{}}
\textit{\$Bankia sigue el crack bursátil. -1,925 euros, del IBEX35 \#mayorcaída https://t.co/S73BxUSKiR}
\end{tabular}\\
& \begin{tabular}[c]{@{}p{14cm}@{}}
`\$Bankia follows the stock market crash. -1,925 euros, the biggest fall in IBEX35 \#biggestdrop https://t.co/S73BxUSKiR'
\end{tabular}\\\hline
\multirow{2}{*}{\bf After} & \begin{tabular}[c]{@{}p{14cm}@{}}
\textit{\textsc{ticker}\#1 seguir bursátil \textsc{negative} euros \textsc{ticker}\#2 mayor caída}
\end{tabular}\\
& \begin{tabular}[c]{@{}p{14cm}@{}}
`\textsc{ticker}\#1 follow stock market \textsc{negative} euros \textsc{ticker}\#2 biggest fall'
\end{tabular}
\\\bottomrule
\end{tabular}
\end{table*}

\subsubsection{Feature engineering}
\label{sec:feature_engineering_architecture}

This module generates and engineers features from the text or from external quantitative data sources related to stock markets and assets. Table \ref{tab:features} summarises the features we consider:

\begin{itemize}

 \item \textbf{Textual features}: char-grams, word-grams and word tokens, plus a bag of words (\textsc{bow}) composed of the most frequent words and bigrams that are unique for each emotion category. 

 \item \textbf{Numerical features}: these features characterise each entry by tweet length; amount of numerical values and percentages (negative, positive and total); amount of financial abbreviations\footref{financial_lexica}, exclamation and interrogative symbols; amount of adverbs (total, negative, positive, expressing doubt and intensifiers); and amount of words with polarity\footnote{Available at \url{https://www.gti.uvigo.es/index.php/en/resources/8-lexicon-of-polarity-and-list-of-emojis-by-polarity-and-emotion-for-application-in-the-financial-field}, January 2023.} (negative, neutral and positive) and describing general emotions\footnote{Available at \url{https://www.cic.ipn.mx/~sidorov/SEL.txt}, January 2023.}. To create these features, we perform morphological and syntactic parsing and exploit our lexica\footref{financial_lexica}.
 
 \item \textbf{Boolean features}: we compute the variations of stock market assets as the differences between their previous and posterior prices by considering the temporary window between the working day before the tweet was posted and the end of the following working day. Then, we indicate if the trend is upwards or downwards.

\end{itemize}

\begin{table*}[ht!]
\centering
\caption{\label{tab:features} Features and target of the Machine Learning models.}
\begin{tabular}{llll}
\toprule
\bf Type & \bf Num. & \bf Feature name & \bf Description \\\hline

\multirow{6}{*}{Textual} & 1 & CHAR\_GRAMS & \begin{tabular}[c]{@{}p{8.5cm}@{}} 
Character $n$-grams\end{tabular}\\

 & 2 & WORD\_GRAMS & \begin{tabular}[c]{@{}p{8.5cm}@{}} 
 Word $n$-grams\end{tabular}\\

 & 3 & WORD\_TOKENS & \begin{tabular}[c]{@{}p{8.5cm}@{}}
 Character $n$-grams only from text inside word boundaries\end{tabular}\\
 
 & 4 & PRE\_BOW & \begin{tabular}[c]{@{}p{8.5cm}@{}} \textsc{bow} of most frequent words and bigrams that are exclusive of precaution emotion\end{tabular}\\
 
 & 5 & NEU\_BOW & \begin{tabular}[c]{@{}p{8.5cm}@{}} \textsc{bow} of most frequent words and bigrams that are exclusive of neutral emotion \end{tabular}\\
 
 & 6 & OPP\_BOW & \begin{tabular}[c]{@{}p{8.5cm}@{}} \textsc{bow} of most frequent words and bigrams that are exclusive of opportunity emotion\end{tabular}\\\hline
 
\multirow{21}{*}{Numerical} & 7 & LEN\_TWEET & \begin{tabular}[c]{@{}p{8.5cm}@{}} Tweet length\end{tabular}\\

 & 8 & NEG\_NUM & \begin{tabular}[c]{@{}p{8.5cm}@{}} Amount of negative numerical values\end{tabular}\\

 & 9 & POS\_NUM & \begin{tabular}[c]{@{}p{8.5cm}@{}} Amount of positive numerical values\end{tabular}\\
 
 & 10 & TOTAL\_NUM & \begin{tabular}[c]{@{}p{8.5cm}@{}} Amount of numerical values\end{tabular}\\
 
 & 11 & NEG\_PERC & \begin{tabular}[c]{@{}p{8.5cm}@{}} Amount of negative percentages\end{tabular}\\
 
 & 12 & POS\_PERC & \begin{tabular}[c]{@{}p{8.5cm}@{}} Amount of positive percentages\end{tabular}\\
 
 & 13 & TOTAL\_PERC & \begin{tabular}[c]{@{}p{8.5cm}@{}} Amount of percentages\end{tabular}\\
 
 & 14 & FIN\_ABBR & \begin{tabular}[c]{@{}p{8.5cm}@{}} Amount of financial abbreviations\end{tabular}\\
 
 & 15 & EXCLAMATION & \begin{tabular}[c]{@{}p{8.5cm}@{}} Amount of exclamation marks\end{tabular}\\
 
 & 16 & INTERROGATION & \begin{tabular}[c]{@{}p{8.5cm}@{}} Amount of interrogation marks\end{tabular}\\
 
 & 17 & ADVERBS & \begin{tabular}[c]{@{}p{8.5cm}@{}} Amount of adverbs\end{tabular}\\
 
 & 18 & ADVERBS\_NEG & \begin{tabular}[c]{@{}p{8.5cm}@{}} Amount of negative adverbs\end{tabular}\\
 
 & 19 & ADVERBS\_POS & \begin{tabular}[c]{@{}p{8.5cm}@{}} Amount of positive adverbs\end{tabular}\\
 
 & 20 & ADVERBS\_DOUBT & \begin{tabular}[c]{@{}p{8.5cm}@{}} Amount of adverbs expressing doubt\end{tabular}\\
 
 & 21 & ADVERBS\_INT & \begin{tabular}[c]{@{}p{8.5cm}@{}} Amount of intensifying adverbs\end{tabular}\\
 
 & 22 & NEG\_POLARITY & \begin{tabular}[c]{@{}p{8.5cm}@{}} Amount of words with negative polarity\end{tabular}\\
 
 & 23 & NEU\_POLARITY & \begin{tabular}[c]{@{}p{8.5cm}@{}} Amount of words with neutral polarity\end{tabular}\\
 
 & 24 & POS\_POLARITY & \begin{tabular}[c]{@{}p{8.5cm}@{}} Amount of words with positive polarity\end{tabular}\\
 
 & 25 & NEG\_EMOTION & \begin{tabular}[c]{@{}p{8.5cm}@{}} Amount of words that express sadness, anger and other negative emotions\end{tabular}\\

 & 26 & POS\_EMOTION & \begin{tabular}[c]{@{}p{8.5cm}@{}} Amount of words that express happiness, surprise and other positive emotions\end{tabular}\\ \hline
 
\multirow{1}{*}{Boolean} & 27 & TREND & \begin{tabular}[c]{@{}p{8.5cm}@{}} Indicates if the asset exhibits an upward or downward trend\end{tabular}\\\hline

\multirow{1}{*}{Target} & 28 & EMOTION & \begin{tabular}[c]{@{}p{8.5cm}@{}} Financial emotion tag\end{tabular}\\

\bottomrule
\end{tabular}
\end{table*}

Table \ref{tab:features_examples} provides some examples of tweets and their corresponding feature values.

\begin{table*}[ht!]
\centering
\caption{\label{tab:features_examples} Examples of tweet textual content and corresponding feature values. Character ``\_" represents blank spaces.}
\begin{tabular}{lccc}
\toprule
& & \bf Tweet sample 1 & \bf Tweet sample 2\\
& & \begin{tabular}[c]{@{}p{6.5cm}@{}}
\textit{30-07-2019 \#Ibex35 -2,48\% sigen llegando resultados llega agosto mucho cuidado con los piratas de guante blancoveremos si es movido o no... https://t.co/3IX2zsNpEC"}

07-30-2019 \#Ibex35 -2.48\% The results keep coming, August arrives, be very careful with the white glove pirates, we'll see if it moves or not... https://t.co/3IX2zsNpEC 
\end{tabular} & 

\begin{tabular}[c]{@{}p{6.5cm}@{}}
\textit{\#IBEX35 La superación, otra vez, del 9375 del índice, aupado posiblemente por una recuperación de la banca, que ha estado muy castigada, podría darle fortaleza para ir a cotas más altas.}

\#IBEX35 Again, the overcoming of the 9375 reference, possibly boosted by a recovery of the banking system, which has been hit hard, could give the index strength to reach higher levels. \end{tabular}\\\cmidrule{2-4}

\multirow{28}{*}{\bf Features} & 1 &[number,stock,...]&[stock,superación,...] \\
& 2 & [numb,umbe,mber,ber\_,... ] & [stoc,tock,ock\_,ckv\_s,...]\\
& 3 &[\_num,numb,umbe,mber,ber\_,...] & [\_sto,stoc,tock,ock\_,...]\\
& 4 & [mucho cuidar] & -\\
& 5 & - & - \\
& 6 & - & [vez number]\\
& 7 &135 &139 \\
& 8 & 0 & 0\\
& 9 & 0 & 1\\
& 10 & 0 & 1 \\
& 11 & 1 & 0\\
& 12 & 0 & 0\\
& 13 & 1 & 0\\
& 14 & 0 & 0\\
& 15 & 0 & 0\\
& 16 & 0 & 0 \\
& 17 & 2 & 2 \\
& 18 & 1& 0\\
& 19 & 0& 0\\
& 20 & 0& 1\\
& 21 & 1& 1\\
& 22 & 1& 0\\
& 23 & 0& 0\\
& 24 & 2& 2\\
& 25 & 0& 0\\
& 26 & 0& 0\\
& 27 & downward& upward\\
& 28 & precaution& opportunity\\\bottomrule

\end{tabular}
\end{table*}

\subsubsection{Feature analysis and selection}
\label{sec:feature_selection_architecture}

We compute the Pearson Correlation Coefficient (Equation \ref{pearson}) to measure the correlations between the features in Table \ref{tab:features}. For any two features $x$, $y$:

\begin{equation}
r_{xy} = \frac{\sum (x_i - \overline{x}) (y_i - \overline{y})}{\sqrt{\sum (x_i - \overline{x})^2} \sqrt{\sum (y_i - \overline{y})^2}}
\label{pearson}
\end{equation}

\noindent
where $r_{xy} \in [-1,1]$. Then, we apply a selection algorithm to identify the features with higher correlation to reduce the original dimensionality of the data space and, thus, the computational requirements of the models.

\subsection{Stream classification module}
\label{sec:classification_stage_architecture}

We designed a multi-class stacking \citep{Malmasi2018} classifier model. It was implemented in streaming (online) mode \citep{Montiel2018}, which seemed adequate given the nature of social data.

\begin{figure*}[!htbp]
 \centering
 \includegraphics[scale=0.5]{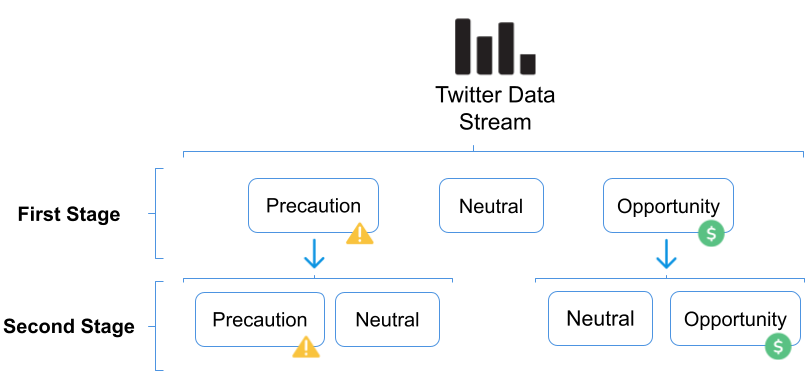}
 \caption{Streaming classification on stacking.}
 \label{fig:stacking}
\end{figure*}

In the stacking ensemble, the first classifier stage differentiates among precaution, opportunity, and neutral (that is, any other) emotions. The second classifier re-evaluates this prediction to improve its accuracy by differentiating again between the corresponding emotion (opportunity or precaution) and the neutral emotion. Figure \ref{fig:stacking} shows the stacking scheme.

We evaluated several Machine Learning algorithms for the stacking ensemble based on their promising performance in similar classification problems \citep{BarronEstrada2020,Kaur2020,Muneer2020}:

\begin{itemize}
 \item Naive Bayes (\textsc{nb})\footnote{Available at \url{https://scikit-multiflow.readthedocs.io/en/latest/api/generated/skmultiflow.bayes.NaiveBayes.html}, January 2023.} \citep{Berrar2019NB}
 \item Decision Tree (\textsc{dt})\footnote{Available at \url{https://scikit-multiflow.readthedocs.io/en/stable/api/generated/skmultiflow.trees.ExtremelyFastDecisionTreeClassifier.html}, January 2023.} \citep{Trabelsi2019} 
 \item Random Forest (\textsc{rf})\footnote{Available at \url{https://scikit-multiflow.readthedocs.io/en/stable/api/generated/skmultiflow.meta.AdaptiveRandomForestClassifier.html}, January 2023.} \citep{Parmar2019}
 \item Stochastic Gradient Descent (\textsc{sgd})\footnote{Available at \url{https://scikit-learn.org/stable/modules/sgd.html}, January 2023.} \citep{Mehlig2021}
\end{itemize}

\label{sec:evaluation_stage}

\section{Evaluation}
\label{sec:results}

All the experiments were executed on a computer with the following specifications:
\begin{itemize}
 \item \textbf{Operating System}: Ubuntu 18.04.2 \textsc{lts} 64 bits
 \item \textbf{Processor}: Intel\@Core i9-9900K 3.60 \textsc{gh}z 
 \item \textbf{RAM}: 32 \textsc{gb} \textsc{ddr}4 
 \item \textbf{Disk}: 500 Gb (7200 rpm \textsc{sata}) + 256 \textsc{gb} \textsc{ssd}
\end{itemize}

\subsection{Experimental data set}

The experimental data set\footnote{Available from the corresponding author on reasonable request.} is composed of 5,000 tweets manually tagged as explained in Section \ref{annseg}. It was collected using the Twitter \textsc{api}\footnote{Available at \url{https://developer.twitter.com/en/docs}, January 2023.} from January 14th to June 21st 2020. It is similar in size to the data sets in other studies \citep{Chatzis2018,Al-Smadi2019,Simester2019,Tuke2020,Plaza-del-Arco2021}. Table \ref{tab:entries_dataset} shows some examples of the texts of the entries.

\subsection{Constituency parsing module}

For constituency parsing, we used the CoreNLP library\footnote{Available at \url{https://stanfordnlp.github.io/CoreNLP}, January 2023.} along with the techniques and considerations explained in Section \ref{sec:constituency_parsing_module_architecture}.

\subsection{Annotation of segments}
\label{annseg}

The constituency parsing module may produce segments with one or several different assets. In the second case, each segment in the training set is replicated as many times as the different assets it contains to ensure a correct annotation. For each replica, only one asset is tagged as \textsc{ticker} while the rest are tagged as \textsc{other\_ticker}. Each replica is annotated as opportunity, precaution or neutral by focusing on the role of \textsc{ticker}.

The segmented entries of the data set were manually labelled by three experts in both \textsc{nlp} and finance. Table \ref{tab:dataset_distribution} shows the resulting distribution of the segments in the experimental data set by emotion category tag. 
 
\begin{table*}[!htbp]
\centering
\caption{Examples of the texts of the entries of the data set.}
\begin{tabular}{lc}
\toprule
\textit{A pesar del difícil año para el sector, SAB.MC mantiene la recomendación de compra para MT}
\\
`Despite the difficult year for the sector, SAB.MC maintains a buy recommendation for MT'
\\

\textit{Queda un recorrido al alza considerable en Caixabank y en BBVA}
\\
`There is a considerable upward path in Caixabank and BBVA'
\\\hline

\textit{Bankia sigue el crack bursátil. 1,925 euros, mayor caída del IBEX35}
\\
`Bankia follows the stock market crash. 1,925 euros, the biggest fall in IBEX35'
\\

\textit{NKE rompe con AMZN}
\\
`NKE breaks with AMZN'
\\\hline

\textit{Así va el IBEX35 en lo que llevamos del mes de agosto}
\\
`This is how IBEX35 has gone so far in August'
\\

\textit{AAPL y KO veremos que nos espera}
\\
`AAPL and KO we will see what awaits us'
\\
\bottomrule
\end{tabular}
\label{tab:entries_dataset}
\end{table*}
 
\begin{table}[!htbp]
\centering
\caption{Distribution of annotated segments in the data set by financial emotion.}
\begin{tabular}{cc}
\toprule \textbf{Emotion} & \textbf{Number of segments} \\ \toprule
Precaution ($P^-$) & 1644 \\
Neutral ($N$) & 4172 \\ 
Opportunity ($O^+$) & 2392\\
\hline
Total & 8208 \\ \bottomrule
\end{tabular}
\label{tab:dataset_distribution}
\end{table}

To assess the consistency of the annotation processes of financial emotions precaution, opportunity and neutral, we calculated Alpha-reliability and accuracy \citep{Giannantonio2010}. Table \ref{coincidenceMatrix} shows the coincidence matrix of the annotators. Note that the elements in the diagonal of the matrix count the tweets on which the three annotators fully agreed. Tables \ref{alphaMatrix} and \ref{accurayMatrix} show the inter-agreement by pairs of annotators. For reference purposes, the literature considers that values above 0.667 are acceptable and that those near 0.8 are optimal, as in our case \citep{Rash2019,Salminen2019,Seite2019,Kilicoglu2021}.

\begin{table}[!htbp]
\centering
\caption{\label{coincidenceMatrix}Coincidence matrix of the annotation process.} 
\begin{tabular}{lccc}
\toprule
\multicolumn{1}{l}{} & \textbf{$P^-$} & \textbf{$N$} & \textbf{$O^+$} \\ \hline
\textbf{$P^-$} & 2827 & 341 & 102\\
\textbf{$N$} & 341 & 7505 & 662\\
\textbf{$O^+$} & 102 & 662 & 3466\\
\bottomrule
\end{tabular}
\end{table}

\begin{table}[!htbp]
\centering
\caption{\label{alphaMatrix}Inter-agreement Alpha-reliability of emotion tag by pairs of annotators.}
\begin{tabular}{lccccc}\toprule
\multicolumn{1}{c}{} & {\bf Ann. 1}& {\bf Ann. 2}& {\bf Ann. 3}\\\hline
{\bf Annotator 1} & - & 0.792 & 0.703\\
{\bf Annotator 2} & 0.792 & - & 0.819\\
{\bf Annotator 3} & 0.703 & 0.819 & -\\\bottomrule
\end{tabular}
\end{table}

\begin{table}[!htbp]
\centering
\caption{\label{accurayMatrix}Inter-agreement accuracy of emotion tag by pairs of annotators.}
\begin{tabular}{lccccc}\toprule
\multicolumn{1}{c}{} & {\bf Ann. 1}& {\bf Ann. 2}& {\bf Ann. 3}\\\hline
{\bf Annotator 1} & - & 0.874 & 0.822\\
{\bf Annotator 2} & 0.874 & - & 0.890\\
{\bf Annotator 3} & 0.822 & 0.890 & -\\\bottomrule
\end{tabular}
\end{table}

\subsection{Data processing module}

Next, we describe the implementations of the text processing, feature engineering, and feature analysis and selection modules.

\subsubsection{Text processing}

Text processing is based on our financial lexica, and the techniques explained in Section \ref{sec:text_processing_architecture}. Particularly, for the lemmatisation process, we used the Freeling library\footnote{Available at \url{http://nlp.lsi.upc.edu/freeling/node/1}, January 2023.}.

\subsubsection{Feature engineering}

Regarding feature engineering, we applied the methodology explained in Section \ref{sec:feature_engineering_architecture}. Table \ref{tab:features} shows the list of features. Specifically, to produce the textual features (1 to 3 in Table \ref{tab:features}) we used \texttt{CountVectorizer}\footnote{Available at \url{https://scikit-learn.org/stable/modules/generated/sklearn.feature_extraction.text.CountVectorizer.html}, January 2023.} and \texttt{GridSearchCV}\footnote{Available at \url{https://scikit-learn.org/stable/modules/generated/sklearn.model\_selection.GridSearchCV.html}, January 2023.}, both from the Scikit-Learn Python library\footnote{Available at \url{https://scikit-learn.org}, January 2023.}, with the parameter ranges in Listing \ref{configuration_parameters}. The optimal parameter settings were \texttt{max\_df = 0.5}, \texttt{min\_df = 0.001} and \texttt{ngram\_range = (1,4)}. Also, for \textsc{bow} features (4 to 6 in Table \ref{tab:features}) we used \texttt{CountVectorizer}, by applying frequency sorting and keeping the 500 most frequent words and bigrams that were exclusive to each emotion category. Finally, asset trends (feature 27 in Table \ref{tab:features}) were obtained from Yahoo! Finance.
 
\begin{lstlisting}[frame=single,caption={Parameter ranges for the generation of $n$-grams.}, label={configuration_parameters}]
max_df: (0.3,0.35,0.4,0.5,0.7,0.8,1)
min_df: (0,0.001,0.005,0.008,0.01)
ngram_range: ((1,1),(1,2),(1,3),(1,4),(1,4),(1,5),(1,6),(1,7))
\end{lstlisting}

\subsubsection{Feature analysis and selection}
\label{feansel}

First, we estimated the contributions of the features to the target using the Pearson correlation coefficient, as explained in Section \ref{sec:feature_selection_architecture}. Table \ref{tab:heatmaps} shows the most correlated features. They correspond to percentage values, polarity lexica and assets' trends. 

Consequently, both financial historical data and knowledge extracted from social media are relevant to predict negative (precautions) and positive (opportunities) emotions. The moderate correlation levels indicate that the classification problem can be addressed with Machine Learning algorithms.

\begin{table}[!ht]
\centering
\caption{\label{tab:heatmaps}Most correlated features with the target (feature identifiers from Table \ref{tab:features}).}
\begin{tabular}{ccc}
\toprule
\bf Feature num. & \bf Correlation\\\hline
24 & 0.10\\
12 & 0.14\\
27 & 0.21\\\bottomrule
\end{tabular}
\end{table}

We then selected the features using the transformer wrapper \texttt{SelectPercentile}\footnote{Available at \url{https://scikit-learn.org/stable/modules/feature\_selection.html}, January 2023.} method from Scikit-Learn with $\chi^2$ score function and 15th percentile threshold, \textit{i.e.}, the features above that threshold were considered relevant. In the end, the features we selected were 5, 8-10, 19, 21, 22 and 24 in Table \ref{tab:features}, plus 1,734 out of 11,555 $n$-gram features.

\subsection{Streaming classification}\label{sec:numericaltest}

In this section, we evaluate the final performance of our system to detect financial opportunities and precautions. The results were computed using two different streaming approaches. The first is a single-stage scheme as a baseline, using the classifiers in Section \ref{sec:classification_stage_architecture}. The second is the multi-class stacking ensemble described in the same section. Since both were implemented in streaming mode, they were progressively tested and trained by sequentially using each sample from the experimental data set to test the model (\textit{i.e.}, to predict) and then to train the model (\textit{i.e.}, for a partial fit). Performance metrics are obtained as their incremental averages. In particular, we employed the \texttt{EvaluatePrequential}\footnote{Available at \url{https://scikit-multiflow.readthedocs.io/en/stable/api/generated/skmultiflow.evaluation.EvaluatePrequential.html}, January 2023.} library.

As shown in Table \ref{tab:basic_classifier}, \textsc{rf} and \textsc{sgd} 
exhibited promising performance by attaining accuracies close to 70\% in preliminary tests. However, the recall of \textsc{rf} in the detection of precaution and opportunity emotions, and the precision of \textsc{sgd} for precaution, were rather poor.

\begin{table*}[!ht]
\centering
\caption{\label{tab:basic_classifier}Performance of diverse single-stage Machine Learning classification models for financial emotions, initial tests.}
\begin{tabular}{cccccccc}
\toprule
\bf \multirow{2}{*}{Classifier} & \bf \multirow{2}{*}{Accuracy} & \multicolumn{3}{c}{\bf Precision} & \multicolumn{3}{c}{\bf Recall}\\
& & $P^-$ & $N$ & $O^+$ & $P^-$ & $N$ & $O^+$\\\hline
 \textsc{nb} & 46.82 & 79.14 & \bf 80.65 & 36.17 & 59.55 & 14.89 & \bf 93.77\\
 \textsc{dt} & 60.23 & 74.06 & 64.43 & 49.64 & 25.54 & 76.16 & 57.60\\
 \textsc{rf} & 71.82 & \bf 80.08 & 71.05 & 69.75 & 48.42 & \bf 88.28 & 59.20\\
 \textsc{sgd} & \bf 75.06 & 67.83 & 79.67 & \bf 71.83 & \bf 67.45 & 80.51 & 70.78\\\bottomrule
\end{tabular}
\end{table*}
 
These results motivated us to build the more complex second approach, a multi-class stacking classification scheme. Tables \ref{tab:rf_confusion} and \ref{tab:sgd_confusion} show the confusion matrix of the single-stage \textsc{rf} and \textsc{sgd} classifiers. We can observe (in bold) that most errors are concentrated around neutral predictions. Thus, the stacking approach was designed to maximise precision when discerning non-neutral from neutral financial emotions. 

In each test, we independently optimised the hyperparameters of the classifiers with \texttt{GridSearchCV}. Listings \ref{rf_hype} and \ref{sgd_hype} respectively show the parameter ranges for the \textsc{rf} and \textsc{sgd} models. 

Table \ref{tab:stacking} shows the results for the two best classifiers, \textsc{rf} and \textsc{sgd}, using the single-stage and stacking approaches. Particularly, row 1 of Table \ref{tab:stacking} contains the results of the single-stage classifiers with the selection of features in Section \ref{feansel}. Hyperparameter optimisation further improved the best classifiers (with a $\sim$10\% precision increase for the \textsc{rf} classifier compared to Table \ref{tab:basic_classifier}\footnote{Results obtained without hyperparameter optimisation.}). By applying the stacking approach (see row 2), accuracy was similar for better precision in the detection of opportunities and precautions, especially with the \textsc{sgd} classifier, at the cost of some degradation of the precision for neutral emotions. Consequently, the recall of neutral emotions improved.

\begin{table}[!ht]
\centering
\caption{\label{tab:rf_confusion}Single-stage \textsc{rf} confusion matrix, without \textsc{bow} features.}
\begin{tabular}{lccc}
\toprule
& $P^-$ & $N$ & $O^+$ \\\hline
$P^-$ & 796 & \textbf{620} & 228 \\
$N$ & \textbf{103} & 3683 & \textbf{386} \\
$O^+$ & 95 & \textbf{881} & 1416 \\\bottomrule
\end{tabular}
\end{table}

\begin{table}[!ht]
\centering
\caption{\label{tab:sgd_confusion}Single-stage \textsc{sgd} confusion matrix, without \textsc{bow} features.}
\begin{tabular}{lccc}
\toprule
& $P^-$ & $N$ & $O^+$ \\\hline
$P^-$ & 1109 & \textbf{342} & 193 \\
$N$ & \textbf{342} & 3359 & \textbf{471} \\
$O^+$ & 184 & \textbf{515} & 1693 \\\bottomrule
\end{tabular}
\end{table}

We further improved the system by incorporating features 4-6 in Table \ref{tab:features}. Rows 3 and 4 of Table \ref{tab:stacking} show the performance of the single-stage and stacking approaches of the experiments in rows 1 and 2, by adding those extra \textsc{bow} features. This modification enhanced the system significantly in general (\textit{i.e.}, 10\% improvement to 80\% in the recall of opportunities with the single-stage \textsc{sgd} classifier, 6\% improvement to 91\% in the precision of precaution with the stacked \textsc{rf} classifier, and a dramatic 18\% improvement in the recall of precaution with the stacked \textsc{rf} classifier, from 56\% to 74\%). Overall, the best method was the stacked \textsc{rf} classifier with additional \textsc{bow} features, except for the precision of neutral emotions, for which the single-stage \textsc{sgd} classifier was superior (although its performance was comparable to that of the stacked \textsc{sgd} classifier).

\begin{lstlisting}[frame=single,caption={\textsc{rf} hyperparameter configuration.},label={rf_hype}]
estimators: (10,35,50,100)
max_features: (auto,35,50,100)
lambda: (6,35,50,100)
\end{lstlisting}

\begin{lstlisting}[frame=single,caption={\textsc{sgd} hyperparameter configuration.},label={sgd_hype}]
penalty: (l1,l2,elasticnet)
l1range: (0.05,0.15,0.9)
alpha: (0.001,0.0001,0.00001)
max_iter: (100,1000,10000)
tol: (1e-1,1e-3,1e-5)
\end{lstlisting}

\begin{table*}[!ht]
\centering
\caption{\label{tab:stacking} Performance of \textsc{rf} and \textsc{sgd} single-stage and stacking models in streaming mode.}
\begin{tabular}{lcccccccc}
\toprule
\multicolumn{1}{c}{\bf Configuration} & {\bf Model} & {\bf Accuracy} & \multicolumn{3}{c}{\bf Precision} & \multicolumn{3}{c}{\bf Recall}\\
& & & $P^-$ & $N$ & $O^+$ & $P^-$ & $N$ & $O^+$\\\hline
\multirow{2}{*}{Single-stage (1)} & \textsc{rf} & 77.68 & 84.24 & 75.36 & 79.93 & 57.24 & 91.99 & 66.76\\
& \textsc{sgd} & 76.05 & 70.77 & 79.58 & 73.02 & 67.15 & 82.55 & 70.82\\\hline
\multirow{2}{*}{Stacking (2)} & \textsc{rf} & 77.27 & 85.73 & 73.09 & 84.97 & 55.54 & 94.49 & 62.17\\
& \textsc{sgd} & 76.82 & 76.27 & 76.68 & 77.49 & 62.96 & 87.78 & 67.22\\\hline\hline
Single-stage plus & \textsc{rf} & 84.49 & 90.39 & 81.73 & 86.97 & 74.39 & 93.39 & 75.92\\
{\scriptsize BOW} features (3) & \textsc{sgd} & 83.47 & 79.82 & \textbf{86.07} & 81.38 & \textbf{80.11} & 86.77 & \textbf{80.02}\\\hline
\textbf{Stacking plus} & \textsc{rf} & \bf 84.54 & \textbf{91.43} & 80.45 & \textbf{90.24} & 73.97 & \textbf{95.28} & 73.08\\
\textbf{{\scriptsize BOW} features (4)} & \textsc{sgd} & 84.33 & 84.34 & 84.08 & 84.85 & 78.29 & 90.48 & 77.76\\
\bottomrule
\end{tabular}
\end{table*}

Summing up, the $\sim$70\% accuracy baseline by the \textsc{rf} and \textsc{sgd} models was overcome by the use of \textsc{bow} features (features 4-6 in Table 2). Note the 6\% and 18\% improvements with the stacked \textsc{rf} classifier to reach 91\% and 74\% in precaution precision and recall, respectively. Consequently, we consider the \textsc{rf} classifier the best choice in our scenario, attaining around 85\% accuracy, over 90\% precision and around 75\% recall both for financial precautions and opportunities. These promising results endorse the use of the system to detect financial emotions on Twitter about specific stock market assets as well as its practical interest in decision-making.

The single-stage and stacking results with \textsc{bow} features (rows 3 and 4 in Table \ref{tab:stacking}) reflect the importance of the latter. The performance increase can be explained by the fact that under-represented textual features (word $n$-grams and bigrams) are discarded because their document frequency is strictly lower than the given threshold (see \texttt{min\_df} hyperparameters in Listing \ref{configuration_parameters}), prior to model training. \textsc{bow} features, thanks to the potent corpus they provide, allow generating direct and effective prediction rules for otherwise undetectable patterns. They include word $n$-grams with a high semantic load. A manual inspection of the \textsc{bow} corpus reveals meaningful bigram examples such as \textit{ser bajista} `be bearish', \textit{TICKER quiebra} `TICKER bankruptcy' and \textit{abajo sin} `down without' with respective frequencies of appearance 0.00098, 0.00098 and 0.00085, for financial precaution entries; and \textit{experimentar espectacular} `experience spectacular', \textit{mayor ganancia} `highest profit' and \textit{alcista TICKER} `bullish TICKER', with the same respective frequencies of appearance for financial opportunity entries. If \textsc{bow} features are not forced, they would be excluded from training because their frequency of appearance is less than 0.001\footnote{Note that \texttt{min\_df} = 0.001 in Listing \ref{configuration_parameters} after hyperparameter optimisation.}.

Another relevant aspect to consider to explain the better performance of our approach, including stacking and \textsc{bow} features (see row 4 in Table \ref{tab:stacking}) is the handling of neutral tweets. The bold entries in the confusion matrices in tables \ref{tab:rf_confusion_BOW} and \ref{tab:sgd_confusion_BOW}, in comparison with the respective values in tables \ref{tab:rf_confusion} and \ref{tab:sgd_confusion}, show the reduction of categorisation errors due to misclassified neutral tweets as precautions and opportunities, and the other way round (see tables \ref{tab:comparison_rf} and \ref{tab:comparison_sgd}). These are the hardest to avoid in principle since mutually misclassified precautions and opportunities are much less frequent. In particular, with these modifications, there is 67\% and 33\% prediction improvement for neutral entries that were previously incorrectly predicted as opportunities and precautions, respectively (see Table \ref{tab:comparison_rf}).

\begin{table}[!ht]
\centering
\caption{\label{tab:rf_confusion_BOW} Stacking plus \textsc{bow} features \textsc{rf} confusion matrix.}
\begin{tabular}{lccc}
\toprule
& $P^-$ & $N$ & $O^+$ \\\hline
$P^-$ & 1216 & \textbf{367} & 61 \\
$N$ & \textbf{69} & 3975 & \textbf{128} \\
$O^+$ & 45 & \textbf{599} & 1748 \\\bottomrule
\end{tabular}
\end{table}

\begin{table}[!ht]
\centering
\caption{\label{tab:sgd_confusion_BOW} Stacking plus \textsc{bow} features \textsc{sgd} confusion matrix.}
\begin{tabular}{lccc}
\toprule
& $P^-$ & $N$ & $O^+$ \\\hline
$P^-$ & 1287 & \textbf{268} & 89 \\
$N$ & \textbf{154} & 3775 & \textbf{243} \\
$O^+$ & 85 & \textbf{447} & 1860 \\\bottomrule
\end{tabular}
\end{table}

\begin{table}[!ht]
\centering
\caption{\label{tab:comparison_rf} Relative prediction improvement (percentage) with stacking plus \textsc{bow} features compared to single-stage, \textsc{rf} model.}
\begin{tabular}{lccc}
\toprule
& $P^-$ & $N$ & $O^+$ \\\hline
$P^-$ & - & -40.81\% & -73.25\% \\
$N$ & -33.01\% & - & -66.84\% \\
$O^+$ & -52.63\% & -32.01\% & - \\\bottomrule
\end{tabular}
\end{table}

\begin{table}[!ht]
\centering
\caption{\label{tab:comparison_sgd} Relative prediction improvement (percentage) with stacking plus \textsc{bow} features compared to single-stage, \textsc{sgd} model.}
\begin{tabular}{lccc}
\toprule
& $P^-$ & $N$ & $O^+$ \\\hline
$P^-$ & - & -21.64\% & -53.89\% \\
$N$ & -54.97\% & - & -48.41\% \\
$O^+$ & -53.80\% & -13.20\% & - \\\bottomrule
\end{tabular}
\end{table}

Besides, the combined introduction of stacking and \textsc{bow} features, compared to the single-stage approach (row 1 in Table \ref{tab:stacking}), further reduces the misclassification of precautions and opportunities up to 73\% for the case of precautions incorrectly classified as opportunities with the \textsc{rf} model, 53\% in the opposite case. The rationale behind the stacking approach is that its structure insists on the differentiation between neutral and non-neutral emotions. The latter entries are typically expressed more assertively, so it makes sense to expect that there will be low mutual overlapping between precautions and opportunities. Logically, the entries that are separated in the first stage as neutral may include misclassified precaution and opportunity entries, but this is not relevant if the precision of precautions and opportunities is high at the output of the second stage. Consequently, the stacking approach also seems crucial to improve precision, which is the most relevant performance metric from an application perspective in financial investment decision-making (as wrong suggestions are much more harmful than missing suggestions \citep{Akhtar2019}, \textit{i.e.}, offering fewer recommendations can be acceptable if they are truthful). This effect is well illustrated by the increased precautions precision for the \textsc{sgd} algorithm with stacking by comparing rows 3 and 4 of Table \ref{tab:stacking} (from 79.82\% to 84.34\% for precautions).

\begin{figure*}[!htbp]
 \centering
 \includegraphics[scale=0.35]{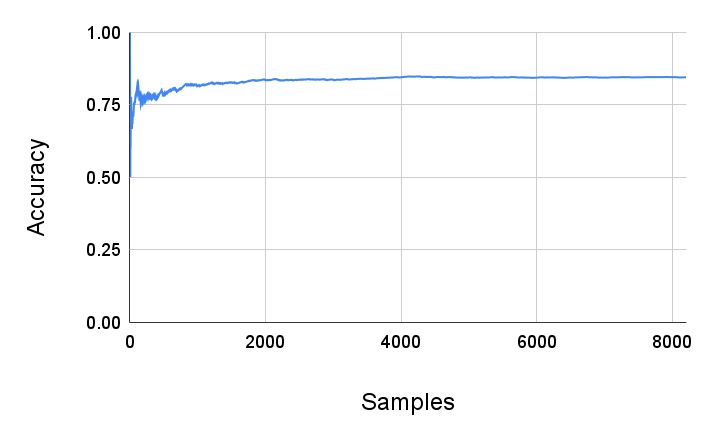}
 \caption{\label{fig:rf_acc_evolution} Evolution of the accuracy of the stacked \textsc{rf} streaming classifier.}
\end{figure*}

Finally, Figure \ref{fig:rf_acc_evolution} shows the evolution over time of the accuracy of the stacked \textsc{rf} streaming classifier, whose cold start ends by the thousandth sample. The performance level is satisfactory to produce usable indicators even for inexpert users. Figure \ref{fig:app} represents a possible view of how these indicators would be used in an investment application. In this example, each entry consists of a text segment, its tickers, and an icon symbolising predicted investment emotion, \includegraphics[height=0.15in]{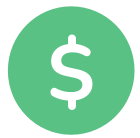} for opportunity and \includegraphics[height=0.15in]{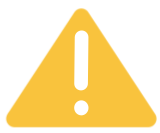} for precaution.

\section{Conclusions}\label{sec:conclusions}

Microblogging platforms such as Twitter provide valuable information for market screening and financial models. They often carry tractable knowledge on investment options that either influence the market or react to it in real-time. Quite often, they include positive and negative educated forecasts, which we term opportunities and precautions in our emotion analysis. 

Motivated by these facts and the interest in helping investors in their decision processes, we propose a novel \textsc{tabea} system that combines \textsc{nlp} techniques and Machine Learning algorithms to predict financial emotions. More in detail, we rely on sophisticated linguistic features, such as sentiment and emotion lexica, along with quantitative financial data and specialised bags of words, and apply them in streaming Machine Learning stacked classification models that process a continuous flow of tweets on-the-fly. This approach is a novel contribution to this work.

\begin{figure*}[!htbp]
 \centering
 \includegraphics[scale=0.18]{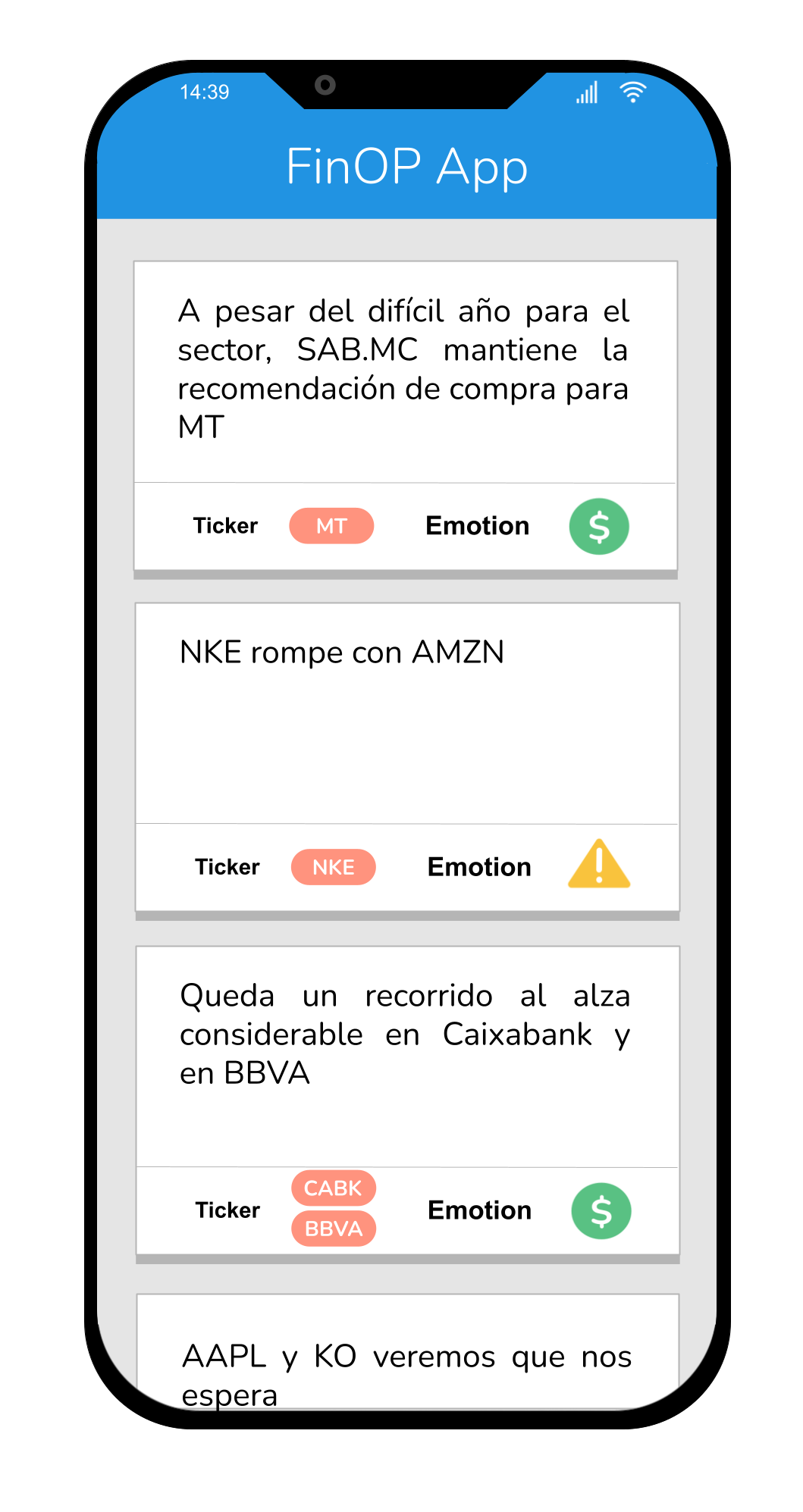}
 \caption{Possible integration of the system in a mobile application.}
 \label{fig:app}
\end{figure*}

The final system achieves over 90\% precision when discerning financial opportunities and precautions on Twitter. The \textsc{tabea} approach considers relevant text segments by focusing on specific assets and provides insightful indicators, as illustrated in Figure \ref{fig:app}.

In future work, we plan to extend our system to a multilingual framework, including English, taking into account that financial jargon is rather technical and very similar in Spanish and other languages, so language-dependent aspects should be easy to handle. We will also further analyse neutrality and ambivalence handling, as well as the potential of our architecture for automatic explainability since the \textsc{rf} algorithms we found so successful are intrinsically adequate to support natural language descriptions of the inner mechanisms of their predictions.

\section*{CRediT authorship contribution statement}

\textbf{Silvia García-Méndez}: Conceptualisation, Methodology, Software, Validation, Formal Analysis, Investigation, Data Curation, Writing - original draft. \textbf{Francisco de Arriba-Pérez}: Conceptualisation, Methodology, Software, Validation, Formal Analysis, Investigation, Data Curation, Writing - original draft. \textbf{Ana Barros-Vila}: Software, Validation, Investigation, Resources, Data Curation, Writing - review \& editing. \textbf{Francisco J. González-Castaño}: Conceptualisation, Methodology, Data Curation, Writing - review \& editing, Supervision.

\section*{Declaration of competing interest}

The authors declare no competing financial interests or personal relationships that could influence the work reported in this paper.

\section*{Acknowledgements}

This work was partially supported by Xunta de Galicia grants ED481B-2021-118 and ED481B-2022-093, Spain; and University of Vigo/CISUG for open access charges.

\bibliography{bibliography}

\end{document}